\documentclass[prl,twocolumn,showpacs,preprintnumbers,amsmath,amssymb,superscriptaddress]{revtex4-1}

\usepackage{amsmath,amssymb,amsfonts,mathrsfs} 	
\usepackage{graphicx}
\usepackage{subfigure}
\usepackage{xcolor}
\usepackage[normalem]{ulem}
\usepackage[colorlinks=true,linkcolor=blue,citecolor=blue,urlcolor=black]{hyperref}

\renewcommand{\[}{\begin{equation}}
\renewcommand{\]}{\end{equation}}
\def\bea{\begin{eqnarray}}
\def\eea{\end{eqnarray}}
\def\nn{\nonumber\\}

\newcommand{\intr}{\int d{\bf r} \;}
\newcommand{\intrc}{\int_{\rm cell} d{\bf r} \;}

\newcommand{\CP}{{\cal P}}

\renewcommand{\k}{{\bf k}}

\newcommand{\vc}{V_{\rm cell}}
\renewcommand{\r}{{\bf r}}
\newcommand{\R}{{\bf R}}

\newcommand{\lt}{ \mathcal{F}_{\alpha\beta}}
\newcommand{\lm}{ \mathcal{L}_{\alpha\beta}}
\newcommand{\tlm}{ \tilde{\mathcal{L}}_{\alpha\beta}}

\newcommand{\ylm}{\tilde{\mathcal{L}}_{yy}}
\newcommand{\equ}[1]{Eq.~(\ref{#1})}

\def\bra#1{\langle#1\vert}
\def\ket#1{\vert#1\rangle}
\def\ev#1{\langle#1\rangle}
\def\me#1#2#3{\langle#1| \, #2 \, |#3\rangle}
\def\runtime{(\the\time)\qquad\the\month/\the\day/\the\year}
\def\today
 {\count10=\year\advance\count10 by -2000 \number\day--\ifcase
  \month \or Jan\or Feb\or Mar\or Apr\or May\or Jun\or
             Jul\or Aug\or Sep\or Oct\or Nov\or Dec\fi--\number\count10}

\def\hour{\count10=\time\count11=\count10
\divide\count10 by 60 \count12=\count10
\multiply\count12 by 60 \advance\count11 by -\count12\count12=0
\number\count10 :\ifnum\count11 < 10 \number\count12\fi\number\count11}

\begin{document}

\title{Local theory of the insulating state}

\author{Antimo Marrazzo}
\email{antimo.marrazzo@epfl.ch}
\affiliation{Theory and Simulation of Materials (THEOS) and National Centre for Computational Design and Discovery of Novel Materials (MARVEL), \'Ecole Polytechnique F\'ed\'erale de Lausanne, 1015 Lausanne, Switzerland}
\author{Raffaele Resta}
\email{resta@democritos.it}
\affiliation{Istituto Officina dei Materiali IOM-CNR, Strada Costiera 11, 314151 Trieste, Italy}
\affiliation{Donostia International Physics Center, 20018 San Sebasti{\'a}n, Spain}

\date{\today}

\begin{abstract}
An insulator differs from a metal because of a different organization of the electrons in their {\it ground state}. In recent years this feature has been probed by means of a geometrical property: the quantum metric tensor, which addresses the system as a whole, and is therefore limited to macroscopically homogenous samples. Here we show that an analogous approach leads to a localization marker, which can detect the metallic vs. insulating character of a given sample region using as sole ingredient the ground state electron distribution, even in the Anderson case (where the spectrum is gapless).
When applied to an insulator with nonzero Chern invariant, our marker is capable of discriminating the insulating nature of the bulk from the conducting nature of the boundary.
Simulations (both model-Hamiltonian and first-principle) on several test cases validate our theory.
\end{abstract}



\maketitle 

The difference between an insulating material and a conducting one is commonly attributed either to spectral properties of the system or to localization properties of the electronic states at the Fermi level (in a mean-field framework). A paradigm change occurred in 1964, when W. Kohn defined the insulating state making neither reference to electronic excitations nor to Fermi-level properties \cite{Kohn64,Kohn68}: the qualitative difference between insulators and conductors manifests itself also in a different organization of the electrons in their many-body {\it ground state}. A series of more recent papers \cite{rap107,Souza00,rap132,rap_a33} has established Kohn's pioneering viewpoint on a sound formal and computational basis, rooted in geometrical concepts. These developments followed (and were inspired by) the modern theory of polarization, based on a Berry phase \cite{Vanderbilt}. The theory---as developed so far---addresses only macroscopically homogeneous systems: either crystalline correlated systems \cite{rap107,Wilkens01,Stella11,Tamura14,Elkhatib15}, or independent-electron systems.
In the latter case spectral properties alone cannot qualitatively discriminate Anderson insulators from metals (both are in fact gapless at the Fermi level), while the geometrical theory is very effective for the task \cite{rap143,rap152}.

In this work we limit ourselves to noninteracting electrons, where for inhomogeneous cases (e.g. in heterojunctions) the metallic/insulating character of a given region is usually probed via the local density of states (LDOS). Here we show---by performing simulations over many test cases---that the metallic/insulating character of the  electronic {\it ground state} can be probed locally, even in the Anderson-insulating case, where the LDOS is of no avail.

The modern formulation of Kohn's theory is based on the quantum metric tensor \cite{Provost80}: it is an extensive quantity having the dimensions of a squared length. We address here the metric tensor per unit volume (area in 2$d$, length in 1$d$); for a macroscopically homogeneous sample we indicate this intensive quantity as $\lm$ (Greek subscripts are Cartesian coordinates throughout). In the noninteracting-electron framework all properties of the many-electron ground state are embedded in the ground state projector; for the sake of simplicity, we give the formulation for ``spinless electrons''. For a bounded sample with square-integrable orbitals the projector is \[ \CP = \sum_{\epsilon_j \leq \mu} \ket{\varphi_j} \bra{\varphi_j} , \] where $\mu$ is the Fermi level, $\ket{\varphi_j}$ are the single-particle orbitals, and $\epsilon_j$ the corresponding energies. The quantum metric tensor has the transparent meaning of the second cumulant moment of the position operator, or equivalently of the ground-state fluctuation of the dipole \cite{Souza00,rap132,rap_a33}: \bea \lm &=& \frac{1}{V} (\, \ev{r_\alpha r_\beta} - \ev{r_\alpha} \ev{r_\beta} \, ) \nn &=&  -\frac{1}{V} \intr \me{\r}{\CP \, [r_\alpha , \CP] \, [r_\beta , \CP]}{\r} \label{basic} .
\eea In the large-sample limit $\lm$ is finite in all insulators, and diverges in all metals; simulations and heuristic arguments altogether suggest that for metallic samples the divergence is of the order of the linear dimension of the sample  \cite{rap143,Bendazzoli12}. If the bounded sample is a crystallite, the integrand in the second line of \equ{basic} is {\it lattice-periodical} in the bulk region of the sample. 

Given that the second line of \equ{basic} is (minus) the trace of the operator $\CP \, [r_\alpha , \CP] \, [r_\beta , \CP]$, divided by the sample volume, we address here the issue of whether the insulating/metallic organization of the electrons in the ground state (in Kohn's words) can be probed by evaluating the trace per unit volume {\it locally} i.e. by integrating the local function \[ \lt(\r) = - \me{\r}{\CP \, [r_\alpha , \CP] \, [r_\beta , \CP]}{\r} \label{lt} \] over a small region in the bulk of the sample. For a homogeneous bounded crystallite we therefore are going to replace $\lm$, \equ{basic}, with its {\it local} counterpart, i.e. \[ \tlm = \frac{1}{V_{\rm cell}}\intrc \lt(\r), \label{ltc} \] where the cell is chosen at the crystallite center. An analogous approach is adopted for either the disordered cases (where the central cell is replaced by a larger region) and for inhomogeneous cases (where the cell is chosen in the appropriate region). 
The main object of the present work is the real symmetric part of $\tlm$, which we are going to name \emph{localization marker}.

We start with 1$d$ bounded chains, by adopting a tight-binding nearest-neighbor Hamiltonian. In the crystalline two-band case the chain is either insulating or metallic according to whether the Fermi level lies in the gap or across a band; in the disordered case the spectrum is gapless but the chain is always Anderson-insulating \cite{Abrahams79}. We adopt the same Hamiltonian as in Ref. \cite{rap143}, where the metric tensor ${\cal L} = {\cal L}_{xx}$, \equ{basic}, has been addressed;
as shown therein, ${\cal L}$ diverges in metallic chains while it converges---to very different values---in the band-insulating and Anderson-insulating cases.

\begin{figure}[t]
\centering
\includegraphics[width=0.8\linewidth]{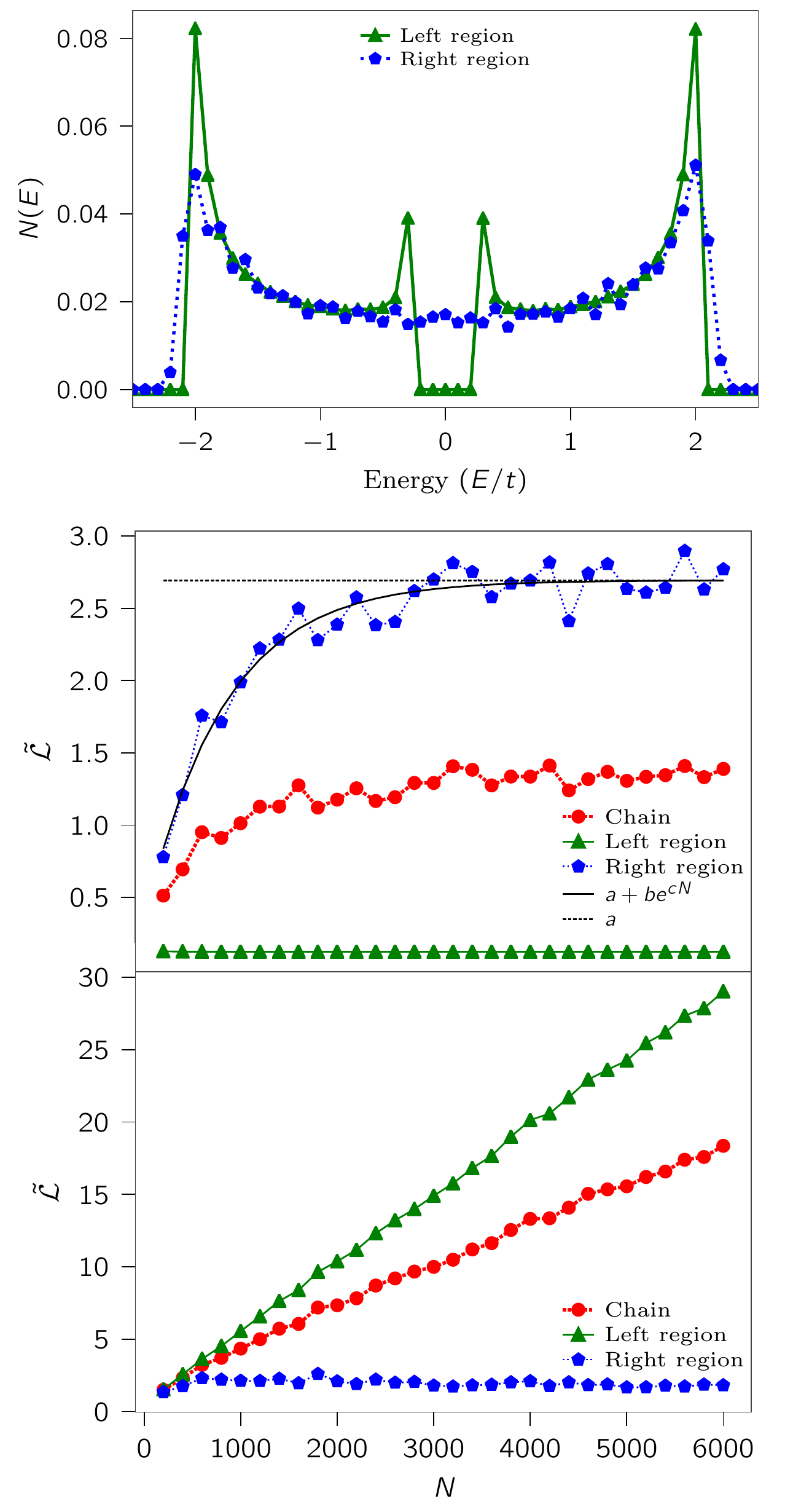} 
\caption{(color online). Results for 1$d$ heterojunctions. Top panel: LDOS for a chain which is crystalline in the left half, and disordered in the right half. Middle panel: $\tilde{\cal L}$ marker for a $\mu=0$ chain (band-insulating in the left half, and Anderson-insulating in the right half). Botton panel: $\tilde{\cal L}$ marker for a $\mu=-1$ chain (metallic in the left half, and Anderson-insulating in the right half).
}
\label{fig:oned} \end{figure}

We have performed simulations over 1$d$ ``heterojunctions'' of up to 6,000 sites, made of two homogenous half-chains, in all the possible combinations of metal, band insulator, and Anderson insulator; the most significant results are displayed in Fig. \ref{fig:oned} \cite{SM_tb}. The top panel shows the LDOS (crystalline vs. disordered), very similar to the global density of states published in Ref. \cite{rap143} (gapped vs. gapless). This LDOS implies that by setting $\mu=0$ the left and right half-chains are band-insulating and Anderson-insulating, respectively, while by setting $\mu=-1$ the left and right half-chains are metallic and Anderson-insulating, respectively. In both cases the LDOS cannnot discriminate correctly, while $\tilde{\cal L}$ accomplishes the task; the metric tensor ${\cal L}$, also shown, yields a kind of average over the whole chain.

\begin{figure}[t]
\centering
\includegraphics[width=0.8\columnwidth]{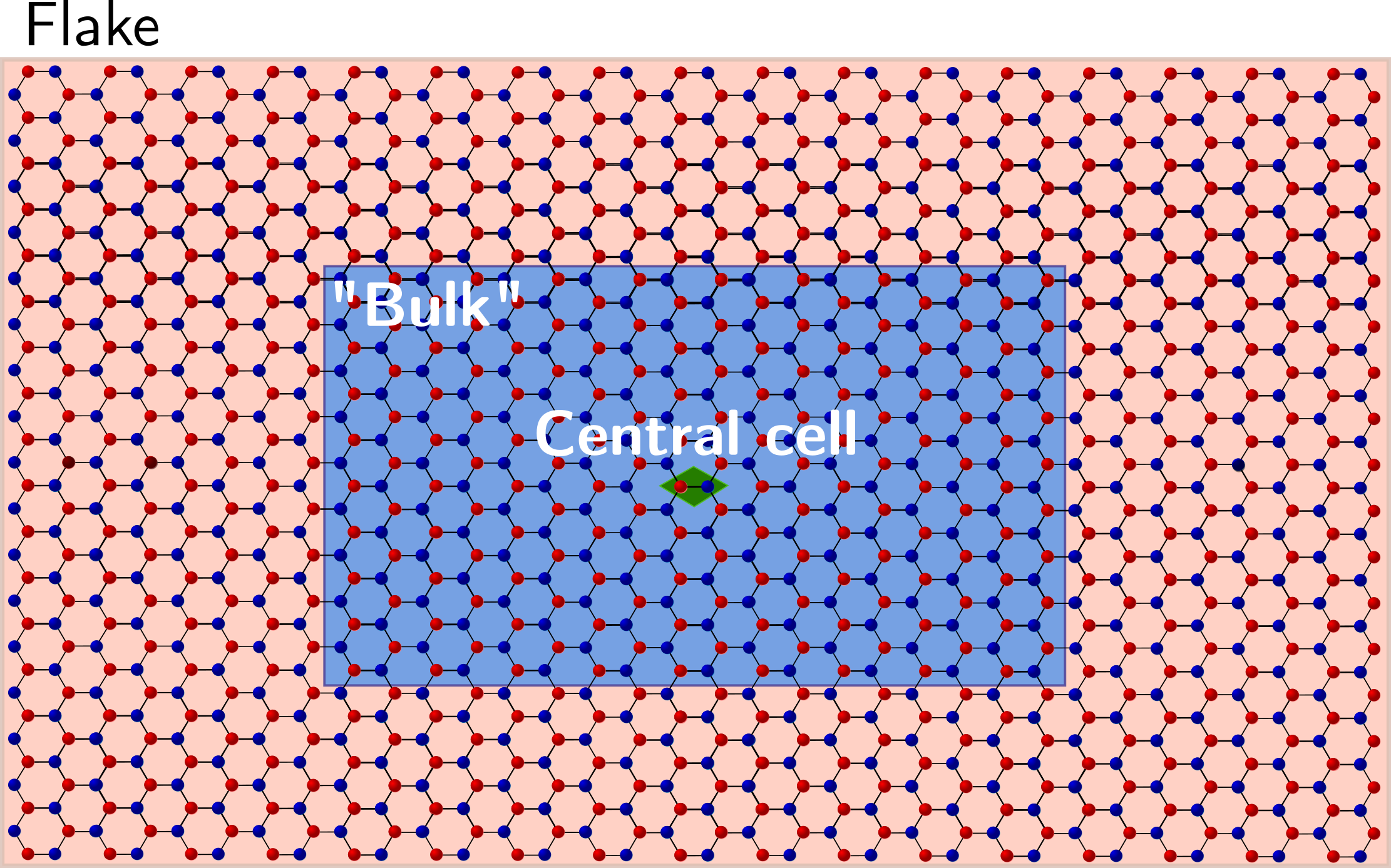}
\caption{(color online). A typical flake (2$d$ crystallite). We have considered flakes with up to 8190 sites, all with the same aspect ratio; the one shown here has 1806 sites. The localization marker $\tlm$ is evaluated either on the central cell (two sites) or by means of analogous integrals on the ``bulk'' region (1/4 of the sites).}
\label{fig:flake} \end{figure}

Next we switch to 2$d$ simulations with model tight-binding Hamiltonians on a honeycomb lattice with two sites per primitive
cell \cite{SM_tb}; a typical flake is displayed in Fig. \ref{fig:flake}.
The electronic structure is described by the orthonormal basis set $\ket{\chi_{\R_\ell}}$, where $\R_\ell$ is a site index. The ground-state projector, \equ{basic}, assumes then the general form \[ \CP = \sum_{\R_\ell{\R_{m}}} P(\R_\ell,\R_m) \; \ket{\chi_{\R_\ell}} \bra{\chi_{\R_m}} . \label{form} \] 

\begin{figure}[t]
\centering
\includegraphics[width=0.8\linewidth]{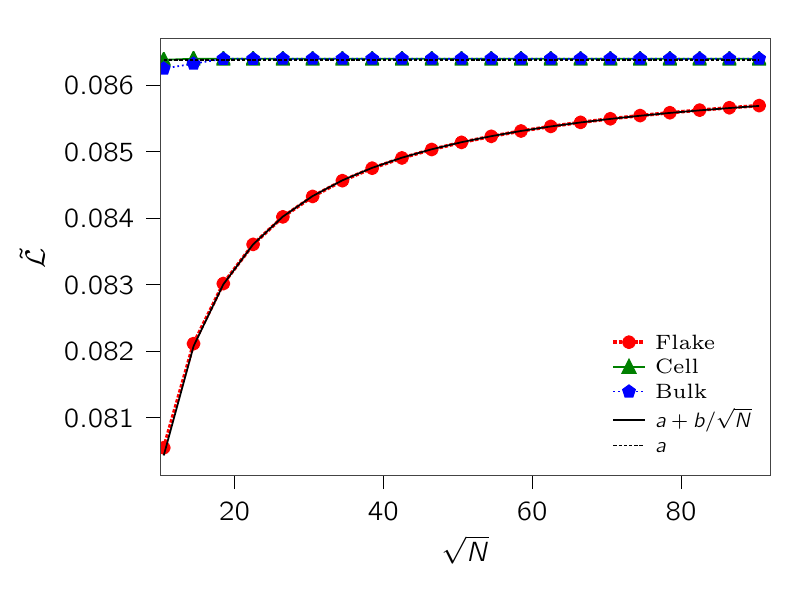}
\caption{(color online). Half-filling homogenous crystalline flake. Cartesian trace of the localization tensor $\lm$, \equ{basic} (labeled ``Flake''), of our localization marker $\tlm$ (labeled ``Cell''), and an analogous formula evaluated over the ``bulk region'' (labeled ``Bulk''), as a function of the flake size. \label{fig:marker1}} \end{figure}

We start with the validation of
our local theory in the simplest cases, where the trace per unit volume of \equ{lt} clearly discriminates the metallic vs. insulating regions and provides indeed the same message as the LDOS. We stress once more the conceptual difference: the former approach probes the ground state, while the latter probes the spectrum. 

Some results are provided in detail in the Supplemental Material \cite{SM_tb}; here we only discuss the insulating (half-filling) homogeneous case: Fig. \ref{fig:marker1} shows the Cartesian trace of $\lm$, of $\tlm$, and of an analogous ``bulk'' quantity where the integral in \equ{ltc} is evaluated over $N/4$ sites (see Fig. \ref{fig:flake}), as a function of the flake size. It is remarkable that the total trace, \equ{basic}, converges to the asymptotic quantum metric quite slowly, only like the inverse linear size of the system; the localization marker $\tlm$ converges instead exponentially. In the crystalline metallic case $\tlm$ diverges like the linear size of the flake (Supplemental Fig. 1 \cite{SM_tb}). We have also verified that our marker can probe the metallic vs. insulating character of the different regions of an inhomogenous sample, by addressing a flake cut through the center by a vertical interface \cite{SM_tb}.

We also address test cases where time-reversal invariance is absent and the insulator is topological, having nonzero Chern invariant: we will show that our marker clearly highlights the insulating character of the bulk and the conducting character of the boundary. To this aim we adopt the Haldane Hamiltonian \cite{Haldane88}, for both a crystalline and a disordered flake \cite{SM_tb} in the topological insulating regime.
It is well known that the flake is insulating in its bulk, while there are topologically protected metallic states at the boundary: it is therefore worth investigating how the different versions of the marker---Cartesian traces of $\lm$ and $\tlm$---actually behave. 

\begin{figure}[t]
\centering
\includegraphics[width=0.8\columnwidth]{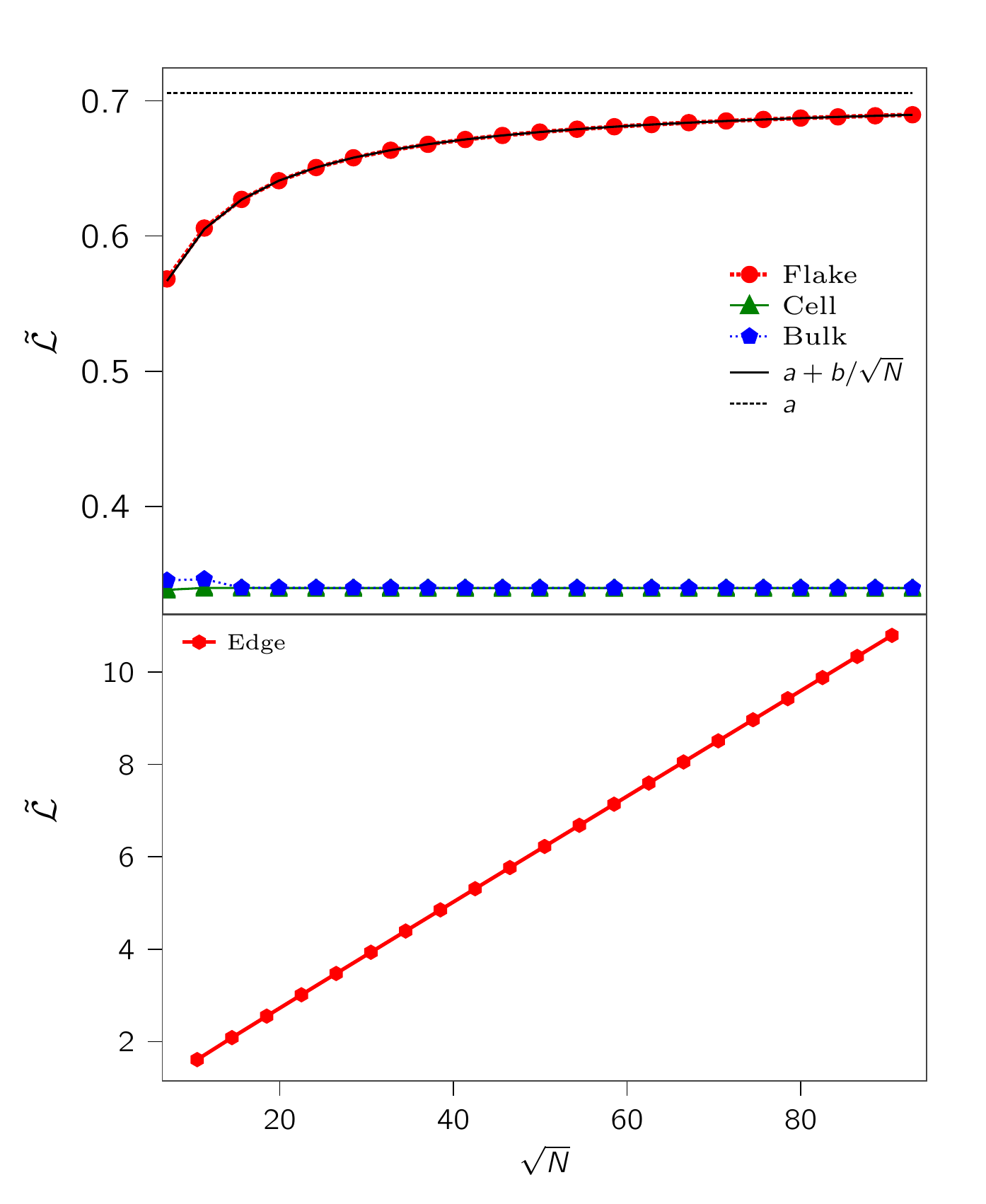}
\caption{(color online). Top panel: Cartesian trace of $\lm$ and of $\tlm$ for a flake cut from a crystalline topological insulator with nonzero Chern number, as a function of the flake size. Labels as in the previous figures. Bottom panel: Cartesian trace of the localization marker $\tlm$, averaged over the boundary cells.
}
\label{fig:chern} \end{figure}

The relevant quantities are plotted in Fig. \ref{fig:chern}. The bottom panel shows that the trace of $\tlm$ diverges like the linear dimension $L$ of the flake when the cell in \equ{ltc} is chosen at the flake boundary (the average over the boundary cells is shown): the boundary is in fact metallic. The top panel shows 
that the trace of $\tlm$ converges fast when the cell is instead chosen in the bulk, and confirms that the bulk is insulating. 

The top panel of Fig. \ref{fig:chern} also shows that the trace of $\lm$ (labelled ``Flake'') converges too, although to a large value. The rationale for the latter feature is that each boundary cell contributes to the integral in \equ{basic} a term proportional to $L$, while the number of boundary cells is also proportional to $L$. The contribution to the total trace is therefore extensive: the trace per unit area is therefore finite (not divergent).

\begin{figure}[t]
\centering
\includegraphics[width=0.8\columnwidth]{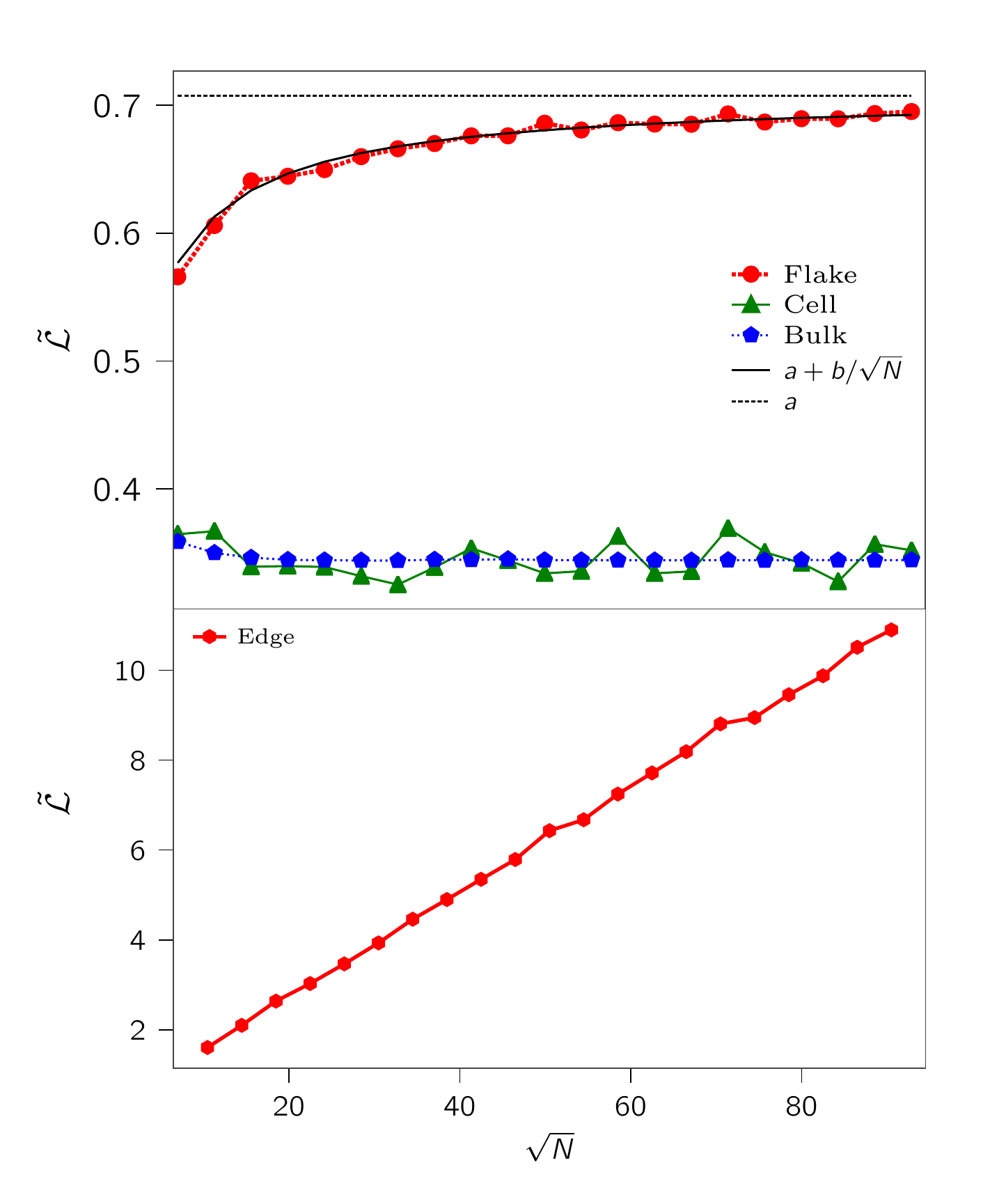}
\caption{(color online). Same as in Fig. \ref{fig:chern} for a topological flake with strong on-site disorder \cite{SM_tb}.}
\label{fig:chern_d} \end{figure}

In the topological case, the insulating behavior is extremely robust with respect to perturbations; here we address the case of strong on-site disorder \cite{SM_tb}. By comparing Fig. \ref{fig:chern} to Fig. \ref{fig:chern_d} it is easily realized that the strong on-site disorder introduces some fluctuations, but does not change at all the key message.

Finally, we are going to present 3$d$ first-principle simulations, {\it not} performed on bounded crystallites; instead, we address a superlattice made of slabs of A and B materials, within periodic boundary conditions (PBCs). To this aim, we rewrite the second line of \equ{basic} as \cite{rap132,rap_a33} \[ \lm = \frac{1}{V}\int d\r \, d\r'\, (\r-\r')_{\alpha}(\r-\r')_{\beta}|\me{\r}{\CP}{\r'}|^2    \label{lt-obc} , \]  which allows switching to an unbounded sample within PBCs.

If the stacking axis is $x$, and A and B are both crystalline materials, then \equ{lt-obc} leads to a localization marker of the form 
\[ \ylm = \frac{1}{\vc} \int_{\vc}  \!\!\!\!\!\! d \r \int d\r' \, (y-y')^2 |\me{\r}{\CP}{\r'}|^2    \label{ylm} , \] where the cell is chosen in the middle of either the A or B regions; the insulating/metallic nature of the slab is then detected by the convergence/divergence of $\ylm$. We have validated \equ{ylm}  by means of PBCs tight-binding simulations, which provided results  equivalent to those  for a bounded flake (Supplemental Fig. 3 \cite{SM_tb}). 

Unfortunately, a first-principle implementation of \equ{ylm} as it stands is computationally prohibitive.  We therefore need a simplified tool, capable of detecting only whether $\ylm$ diverges or converges, without providing its precise value in the insulating cases. 
We adopt the so-called Wannier-interpolation scheme \cite{Marzari12}, which accurately maps the Kohn-Sham Hamiltonian on a tight-binding-like one, capable of describing both insulating and metallic systems. Here we label the basis set as $\ket{\chi_{\R_\ell}}$, where $\R_\ell$ is the orbital center: \[ \R_\ell = \me{\chi_{\R_\ell}}{\r}{\chi_{\R_\ell}} . \label{center} \] With these notations, the projected $\CP$ is identical in form to \equ{form}, where
\[  P({\bf 0}_\ell,\R_m) = \me{\chi_{{\bf 0}_\ell}}{\CP}{\chi_{{\bf R}_m}} , \label{me} \] and the matrix elements are evaluated using a discrete $\k$-point mesh.

If the two basis centers ${\bf 0}_\ell$ and $\R_m$ are both in the middle of a given slab, and distant between themselves in the $y$ (transverse) direction, the qualitative asymptotic behavior of $\me{\r}{\CP}{\r'}$ is reflected into the behavior of the matrix elements  in the $|\R_m - {\bf 0}_\ell| \rightarrow \infty$ limit.
There are several different ways of numerically inspecting asymptotic behaviors. Here---inspired by the tight-binding version of \equ{ylm}---we choose to evaluate the convergence-divergence of the sum \[ \mathscr{L}_{yy} = \frac{1}{\vc} \sum_{{\bf 0}_\ell}  \sum_{\R_m} (0_{\ell y}-R_{my} )^2|P({\bf 0}_\ell,\R_m)|^2    \label{script} . \] We stress that the 
numerical value of $\mathscr{L}_{yy}$, \equ{script}, is {\it different} from the one of  $\ylm$, \equ{ylm}; the key point is that the terms in the summation  
become asymptotically exact when the basis centers are far apart.

Our case study is a periodically repeated (001) supercell of GaAs and Al lattice-matched slabs, with double As termination \cite{SM_dft}. In this geometry the metal and the semiconductor cubic axes are rotated by 45$^o$ around (001), and the lattice-matching condition sets the ratio of
the two cubic lattice constants equal to $1/\sqrt{2}$. Our supercell contains 9 Al layers, 12 Ga layers, and 13 As layers, for a total of 43 atoms (there are two Al atoms per layer). 

\begin{figure}[t]
\centering
\includegraphics[width=\linewidth]{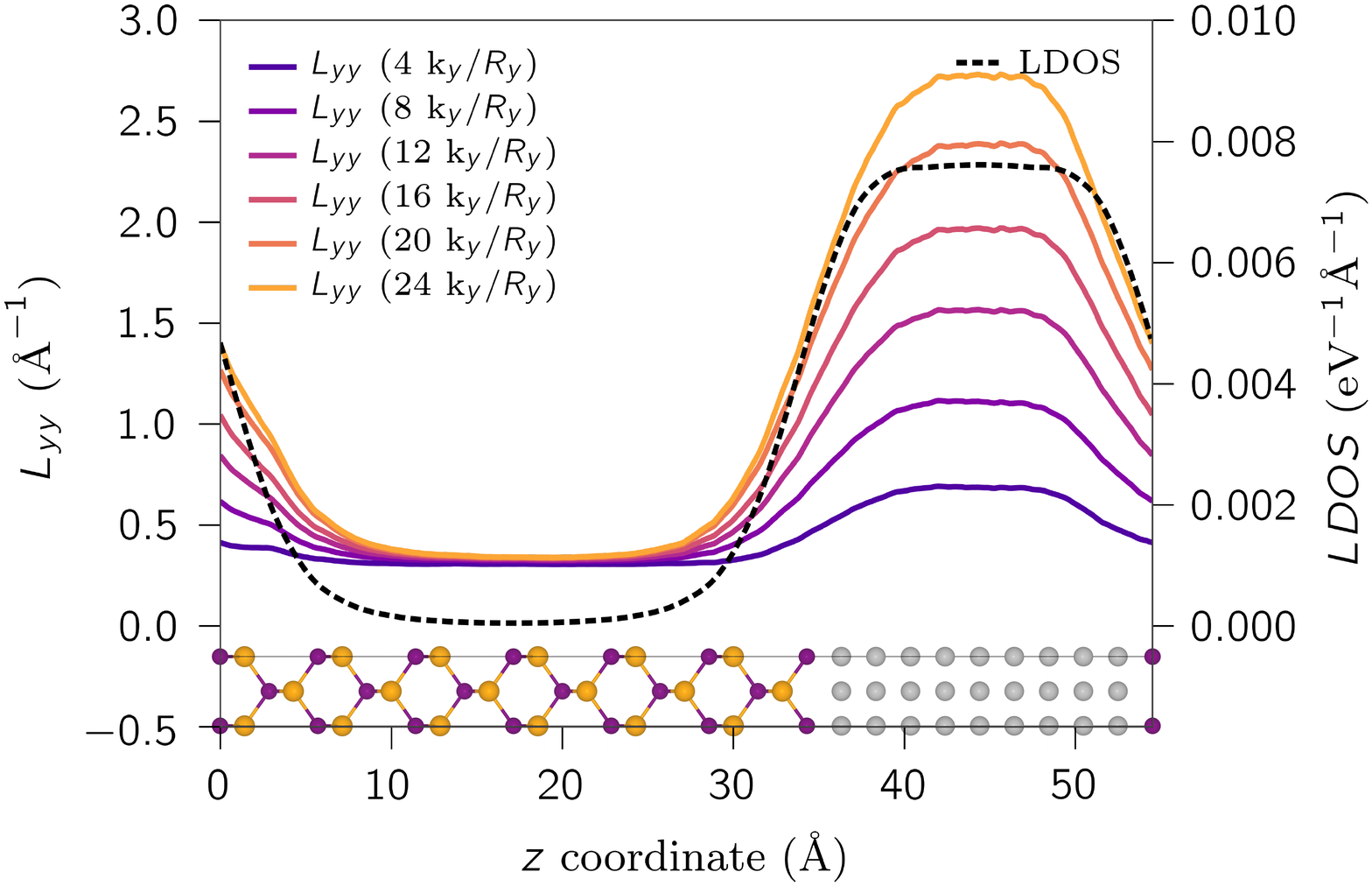}
\caption{(color online) Localization marker $\mathscr{L}_{yy}$ and LDOS at the Fermi level (dashed black line) for a 43-atom AlGaAs heterostructure \cite{SM_dft}. Al atoms are in grey, As in purple and Ga in orange. $\mathscr{L}_{yy}$ is computed using an increasing set of  $\k$ points and $\R$ vectors along the $\hat{y}$ axis (orthogonal to the Al-GaAs axis). In the bulk insulating GaAs region the marker converges very fast to a finite number, due to the exponential decay of the density-matrix, while in the bulk Al it diverges linearly with the number of $\k$ points. All quantities are plotted as double macroscopic averages, defined as in Refs. \cite{rap48,Peressi98}.}
\label{fig:fp-lm} \end{figure}

We show in Fig. \ref{fig:fp-lm}, dashed line,  the LDOS at the Fermi level, filtered with a double macroscopic average \cite{rap48,Peressi98}. As it must be, the LDOS is finite in the metallic region and goes to zero in the insulating region: 
the exponential tail owes to evanescent gap states. 
The novelty of the present work is to show, according to Kohn's viewpoint, that the metallic vs. insulating regions are characterized by a different organization of the electrons in the many-body ground state, without any reference to eigenvalues or spectral properties.

In the GaAs region all the solid lines in Fig. \ref{fig:fp-lm} converge fast to the same value. We remind that our simplified marker $\mathscr{L}_{yy}$ does not provide the same numerical value as the exact localization marker $\ylm$; the finiteness of $\mathscr{L}_{yy}$ proves nonetheless the insulating nature of the ground state electron distribution in the GaAs region.
In the Al region, instead, the different solid lines show the divergence of $\mathscr{L}_{yy}$---ergo of $\ylm$ as well---linear with the number of $\k$ points.

In conclusion, we have shown that the insulating nature of the ground electron distribution can be probed locally, by means of a marker explicitly expressed in terms of the ground state and nothing else, in particular avoiding any reference to either spectral properties or to localization properties of the electronic states at the Fermi level. 
Besides the case of band insulators (model and first-principle), our test cases include Anderson insulators (where the spectral properties are of no avail) and topological insulators (where the bulk is insulating and the boundary is conducting). The simulations presented here address solely independent electrons; nonetheless we argue that our local theory of the insulating state can be extended to correlated electrons as well \cite{suppl}. Our
work paves the way for a unified complete theory of the insulating state, including in principle all kinds of insulators, both homogeneous and heterogeneous (crystallites, heterojunctions, nanostructures), through a localization marker based on the ground-state electronic distribution only.

R.R. acknowledges support by the ONR (USA) Grant No. N00014-17-1-2803; A.M. acknowledges support by the NCCR MARVEL of the Swiss National Science Foundation and the EU Centre of Excellence MaX ``Materials design at the Exascale''.  A.M. acknowledges PRACE for awarding access to computing resources on Marconi at CINECA, Italy and on Piz Daint at CSCS, Switzerland. A.M. also acknowledges useful discussions with Marco Gibertini and Nicola Marzari.

\vfill\vfill


\end{document}